\documentclass[journal]{IEEEtran}
\usepackage{cite}
\usepackage{graphicx}

\ifCLASSINFOpdf
\else
\fi
\usepackage{amsmath,amsfonts,amssymb}
\begin{document}

\title{Mode-walk-off interferometry for position-resolved optical fiber sensing}

\author{Luis Costa,
        Zhongwen Zhan,
        Alireza Marandi

\thanks{This work was supported by the Gordon and Betty Moore Foundation, grant 9500.}
\thanks{Luis Costa (e-mail: luisc@caltech.edu) is with the Department of Electrical Engineering, California Institute of Technology, Pasadena, California 91125, USA and the Seismological Laboratory, California Institute of Technology, Pasadena, California 91125, USA.}% <-this % stops a space
\thanks{ Zhongwen Zhan (e-mail: zwzhan@caltech.edu) is with the Seismological Laboratory, California Institute of Technology, Pasadena, California 91125, USA. Alireza Marandi (e-mail: marandi@caltech.edu) is with the Department of Electrical Engineering, California Institute of Technology, Pasadena, California 91125, USA. (Corresponding author: Alireza Marandi)}% <-this % stops a space
\thanks{Manuscript received XXXX XX, XXXX; revised XXXXX XX, XXXX.}}

\markboth{JOURNAL OF LIGHTWAVE TECHNOLOGY, VOL. XX, NO. XX, XXXXX XX, XXXX}%
{Shell \MakeLowercase{\textit{et al.}}: Bare Demo of IEEEtran.cls for IEEE Journals}

\maketitle

\begin{abstract}
Simultaneously sensing and resolving the position of measurands along an optical fiber enables numerous opportunities, especially for application in environments where massive sensor deployment is not feasible. Despite significant progress in techniques based on round-trip time-of-flight measurements, the need for bidirectional propagation imposes fundamental barriers to their deployment in fiber communication links containing non-reciprocal elements. In this work, we break this barrier by introducing a position-resolved sensing technique based on the interference of two weakly-coupled non-degenerate modes of an optical fiber, as they walk-off through each other. We use this mode-walk-off interferometry to experimentally measure and localize physical changes to the fiber under test (axial strain and temperature) without the typical
requirement of round-trip time-of-flight measurements. The unidirectional propagation requirement of this method makes it compatible with fiber links incorporating non-reciprocal elements, uncovering a path for multiple sensing applications, including ultra-long range distributed sensing in amplified
space-division-multiplexed telecommunication links.
\end{abstract}


% \fi
\IEEEpeerreviewmaketitle

\section{Introduction}

By harnessing intrinsic light-matter interactions, distributed fiber sensing methods enable off-the-shelf optical fibers to operate as highly sensitive sensor arrays \cite{he2021optical-review-das,Lu2019-review}. 
% Such techniques have been demonstrated for measuring a plethora of physical properties, spanning from acoustic \cite{Bashan2018-Impedance,chow2018distributed-Impedance} and optical \cite{Bashan2020-Impedance} properties of the fiber's surrounding environment, to direct measurements of changes to the fiber's physical state, such as pressure \cite{Jia2021-Pressure,ZhangLi-Pressure}, temperature, or axial strain \cite{Zhou2018-StrainTemp,Soriano-Amat2021-StrainTemp}. Distributed acoustic sensing (DAS) \cite{Rao2021-PhaseOTDROFDR,he2021optical-review-das}, in particular,  relates to the subset distributed sensors targeting fast acquisitions of the axial strain profile along a fiber cable, and has historically been applied commercially in fields such as surveillance, pipeline monitoring, and vertical seismic profiling \cite{Rao2021-PhaseOTDROFDR,he2021optical-review-das}.
Multiple variations of these techniques have been used to measure a diverse range of physical parameters, from the acoustic \cite{Bashan2018-Impedance,chow2018distributed-Impedance} and optical \cite{Bashan2020-Impedance} properties of the fiber's surroundings to direct measurements of the interrogated fiber's physical state (e.g., pressure \cite{Jia2021-Pressure,ZhangLi-Pressure}, temperature, or axial strain \cite{Zhou2018-StrainTemp,Soriano-Amat2021-StrainTemp}).
Distributed acoustic sensing (DAS) \cite{Rao2021-PhaseOTDROFDR,he2021optical-review-das} encapsulates the subset of distributed sensors oriented towards fast acquisition of the axial strain profile along a fiber cable, and has traditionally seen application in areas such as surveillance, pipeline monitoring, and vertical seismic profiling \cite{Rao2021-PhaseOTDROFDR,he2021optical-review-das}.

Recently, DAS has captured the attention of the geophysics community for seismic studies \cite{Lior2021-GeoDas,Hammond2019-GeoDas,Williams2019-GeoDas,Fernandez-Ruiz2021-GeoDas}, and considerable research efforts are now aimed at addressing the demands of this new application space. Despite its recent adoption, however, there are plenty of demonstrations of its successful use in geophysical settings: from metropolitan areas \cite{Martins:19-GeoDasDSP} to near-shore deployments \cite{Williams2019-GeoDas}. Nonetheless, environments demanding access to the full extent of ultra-long-haul telecommunication links are still mostly inaccessible through these technologies, which are limited to $\sim$100 km ranges despite remarkable efforts at range extension \cite{Nuno:21-Range,Liu2015a-Range}. 

The deep ocean floor is a particularly relevant example of an environment where traditional geophysical sensing equipment is sparse, costly, and often temporary \cite{Suetsugu2014-OceanbottomTraditionalInstruments}, and where the existence of transoceanic telecommunication fibers presents an exceptional opportunity for fiber-based sensing. So far, attempts at exploring transoceanic fibers for sensing have remained limited in their 
localization capabilities, either unable to discriminate the origin of each strain contribution \cite{MarraScience-OceanFiberSensing,ZhanScience-OceanFiberSensing,Mecozzi:21-OceanFiberSensing}, or able to localize only a few dominant perturbations along the cable through bi-directional measurements \cite{MarraScience-OceanFiberSensing}. %t most localize a few dominant point-like perturbations along the cable through bidirectional interferometry \cite{MarraScience-OceanFiberSensing,EzraIp-Bidirectional}.
Fully distributed DAS techniques, on the other hand, struggle to meet the criteria for sensing in such long-haul cables. The reasons go beyond the aforementioned range limits, as their fundamental reliance on intrinsic backscattering for localization imposes a barrier to massive range enhancement via in-line amplification (owing to the presence of optical isolators in the amplifiers), and the probe pulse's characteristically high peak powers ($\sim$200 mW) render these techniques incompatible with co-propagating data channels \cite{Fernandez-Ruiz2021-GeoDas,Jia:21-Coexistence}, which is especially limiting given the lack of abundant fiber strands in transoceanic cables.

These fundamental roadblocks to ultra-long range DAS motivate the exploration of novel interrogation methods, capable of surpassing the fundamental challenges of current DAS techniques in specific settings. One possible way to expand the design of distributed sensing methods is to target future telecommunication fiber deployments. At the moment, the ability to transport data traffic is lagging under the ever increasing demands of consumers, limited by the gain band of in-line amplifiers and the available power budget of telecommunication fibers \cite{Winzer2020TransmissionSC-SDMTechno}. The current techno-economic landscape suggests that the next iteration of telecommunication fibers will expand capacity through space-division multiplexing (SDM) techniques \cite{Puttnam:21-SDM} (either in the form of few-mode \cite{8004172-FMF} or multicore fibers \cite{Saitoh:16-mcf}) in order to remain economically competitive \cite{dar2018cost-SDMTechno,Winzer2020TransmissionSC-SDMTechno}. Transoceanic distributed sensing methods are expected to closely follow these developments, opening new opportunities to address existing technical limitations by exploring the new spatial degrees of freedom of a multimode telecommunications backbone.

Existing few-mode fiber sensing demonstrations \cite{Li:15-sdmsensrev,9209959-sdmsensrev,Zhao:21-sdmsensrev,kim2021recent-sdmsensrev} have employed the added modes to tackle cross-sensitivity \cite{Li2015FewmodeFM}, improve SNR \cite{8839098}, or in the case of multicore fibers, mostly for complementary measurements using multiple independent channels \cite{9502556,8513845}. One aspect that is less targeted is the need for bidirectional propagation in existing distributed sensing techniques: either for roundtrip time-of-flight measurements \cite{Rao2021-PhaseOTDROFDR,s18041072-PhaseOTDROFDR}, or to engineer a local parametric interaction between two counter-propagating lightwaves \cite{Zhou2018-StrainTemp,Murray:22}. 

Multimode platforms enable the generalization of the time-to-position mapping principles to co-propagating designs, provided that the carried modes/supermodes are weakly coupled and possess different group velocities. This mapping can be observed in reports on the characterization of multimode links and devices \cite{Kono2017-charact,Rommel2017-charact,Fontaine2015-charact,Maruyama2017-charact}, as well as a recent sensing work which proposed the measurement of changes in the local coupling strength as a way to monitor the distribution of transverse stresses applied to the fiber \cite{Jia2021-Pressure,Jia:22}. Yet, more relevant sensing quantities such as strain/temperature remain unexplored, and no sensing demonstration to our knowledge has relied on the baseline coupling and intermode/intercore crosstalk for interrogation - instead, it has been considered as one of the the main drawbacks of few-mode fibers (FMF) for sensing purposes \cite{9209959-sdmsensrev}.

In this work, we demonstrate mode-walk-off interferometry along an optical fiber. This technique enables position-resolved, coherent interrogation of the fiber in a single unidirectional measurement. Through our method, all sections of a weakly-coupled fiber behave as an independent interferometer, whose response is stored at a different time-instant of the temporally broadened output. This enables the recovery of the full profile of common quantities of interest, such as strain and temperature, in a single unidirectional measurement. Our technique employs swept-wavelength interferometry \cite{Fontaine2015-charact, moore2011advances}, while avoiding common drawbacks such as the range limitations originating from the coherence length of widely tunable sources, and the need for a dedicated, optical path-length-matched, local oscillator fiber for recovery of the fiber transmission response. %This design relaxes the coherence length requirements of the source, and the stability requirements between the reference and measurement paths is automatically fulfilled.  Since the sensor will only be sensitive to differences in the optical path between light carried by different modes
% The new trade-offs imposed by our proposed technique are favorable for  its application for geophysics, where the sampling rate can be reduced to a few Hertz and the spatial resolution required for seismic phenomena outside of urban settings is of kilometer lengths (considering seismic wavelengths).
We expect this work to spearhead future unidirectional, multimode distributed sensing designs, able to benefit from in-line amplification and easily integrable with future data-carrying, ultra-long haul telecommunication links, thus unlocking dense (kilometer-long resolutions) strain and temperature sensing in future transoceanic (and other ultra-long-haul), amplified fiber deployments.

\section{Sensing Concept}
\begin{figure*}[hbtp]
\centering
\includegraphics[width=\linewidth]{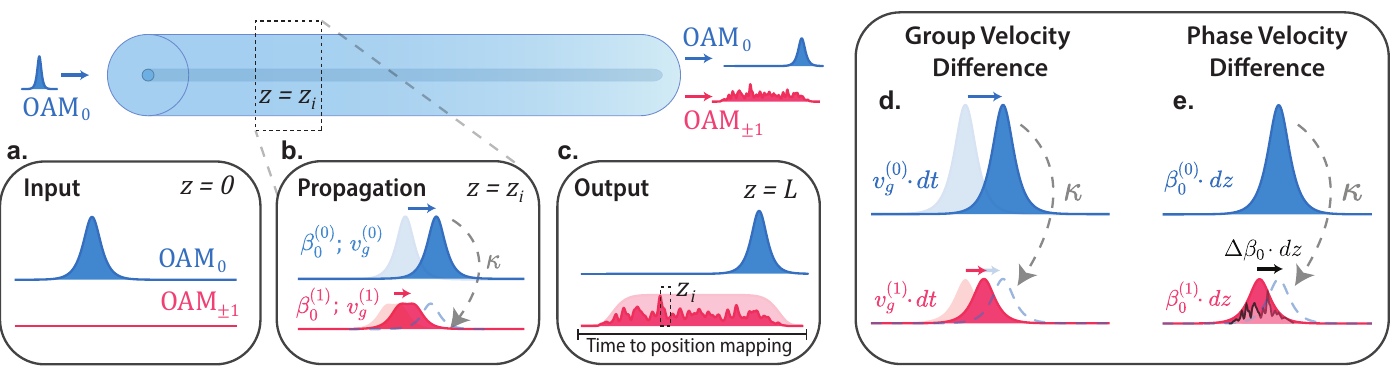}
\caption{Principle of transmission-only distributed sensing. \textbf{a.} Light is launched as the fundamental mode of a fiber carrying at least a pair of non-degenerate modes, of different phase and group velocities. \textbf{b.} Weak distributed coupling (of coupling strength $\kappa$) converts light from the injected mode to higher order modes as it propagates. The difference in group velocities leads to an effective walk-off between previously coupled light and the injected mode beam (\textbf{d.}), such that the point of coupling is mapped to a specific time-instant at the output. \textbf{c.} The difference in group velocity broadens the higher-order mode output, and the difference in phase velocities leads to interference between the newly coupled and previously coupled light  (\textbf{e.}), generating a noise-like broad optical trace at each higher mode output: each time-instant of the OAM\textsubscript{$\pm$ 1} stores the local interference resulting from light traversing two different optical paths.}
\label{fig:Principle}
\end{figure*}

 %, where it is also contrasted with conventional Rayleigh backscattering based methods.
Our transmission-only sensing method is illustrated in figure \ref{fig:Principle}. A short pulse of light is injected into the fundamental mode (OAM\textsubscript{0}, in the Optical Angular Momentum mode basis \cite{Willner:15}) of a two-mode fiber. Under the assumption of constant weak coupling between mode groups most light remains in the same mode as it was injected, but a small fraction is coupled to one of two degenerate higher-order modes (OAM\textsubscript{$\pm$ 1}) at every position in the fiber. 

 After coupling to a higher-order mode, light experiences both a different propagation constant, $\beta^{(\ell)}_0$, and a different group velocity, $v^{(\ell)}_g$ ($\ell$ being the  topological charge of the corresponding OAM mode) for the remaining length of fiber, enabling the estimation of measurand amplitude and the localization of perturbations along the fiber link (see figure \ref{fig:Principle}).

Localization information is encoded in time as a result of the difference in group velocities between mode groups, since light coupled from the fundamental mode at each position will be displaced from pre-existing light in the higher order modes. As a result, at the fiber output, the  OAM\textsubscript{$\pm$ 1} signal will display a broad temporal envelope, each time instant mapping to a specific position in the fiber.

The position information can therefore be recovered via a differential time-of-flight measurement on the higher order mode output. Light coupled to a higher order mode at position $z$ in the fiber will, at the fiber output, exhibit a temporal walk-off relative to light that remained uncoupled proportional to the remaining fiber length ($L_{FUT} - z$) and the differential mode group delay ($ DGD = 1/v_g^{(\pm 1)} - 1/v_g^{(0)}$) between the two mode groups
\begin{equation}
    \Delta t(z) = DGD(L_{FUT} - z).
    \label{eqn:position}
\end{equation}

This method of localization is similar to the traditional backscattering-based methods in conventional distributed techniques, which rely on roundtrip time-of-flight for position discrimination. The differential nature of the measure, however, yields a temporally compressed optical trace (\textit{i.e.}, the obtained impulse response obtained for a given pair of input/output modes) compared to traditional roundtrip  measurements (by a factor $CF \approx -\frac{2 v^{(0)}_g}{ \Delta v_g}$). A narrower optical trace implies a natural penalty to the spatial resolution, but relaxes the fundamental limits of acquisition rate. Few-mode fibers typically have DGDs in the order of a few picoseconds per meter, implying a $CF$ of the order of $10^3$.

Perturbations to the local optical path of the fiber (induced by strain or temperature, in this case) 
can be measured by observing changes to the resulting local interference from light coupling, similar to Rayleigh scattering based systems \cite{Rao2021-PhaseOTDROFDR,s18041072-PhaseOTDROFDR}. As light couples from the OAM\textsubscript{0} mode to any of the higher order OAM\textsubscript{$\pm$ 1} modes, it shall interfere with pre-existing light in the higher order mode with which it overlaps. The optical path difference accumulated due to propagation as different modes over any length of fiber enables us to perceive the fiber as a stack of effective interferometers (sensing points), which we are able to individually access by a time-of-flight measurement to determine the position where coupling happened. The resulting noise-like output of the higher order mode (the optical trace), stores the response of each of the sensing points of the fiber at different time instants.

Recovering the measurand information can then be achieved by either observing the changes to the instantaneous phase evolution along the obtained higher order mode output (analogous to coherent phase-sensitive OTDR interrogation \cite{Wang:16}), or by probing the equivalent frequency shift that compensates the change in the intermodal interference due to a perturbation-induced change in local optical path difference (in analogy to frequency-demodulation phase-sensitive OTDR methods \cite{Lu:20,Koyamada2009}). We opt for the latter approach, since it avoids problems resulting from cumulative measurements of phase, such as poor phase estimation at points of fading and ambiguity due to phase unwrapping errors. 

Consider the interference happening at position $z$ in the fiber, as newly coupled light interferes after travelling a short length $dz$ (stored at a specific time-instant of the recovered optical trace, given by equation \ref{eqn:position}). The phase difference accumulated between the two interfering waves due to the difference in propagation constants will be $\Delta \varphi (z) = \Delta \beta dz$
%PLUS PI/2?? 
 (for $\Delta \beta = \beta^{(0)}_0 - \beta^{(\pm 1)}_0$), which may be re-written as
\begin{equation}
    \Delta \varphi (z) = \frac{2 \pi}{c_0} (\Delta n(z) \cdot dz) \nu_{0}.
    \label{eqn:phase}
\end{equation}

Equation \ref{eqn:phase} illustrates the equivalence of altering the optical path difference $(\Delta n(z) \cdot dz)$ ($\Delta n = n^{(0)} - n^{(1)}$, $n^{(\ell)}$ being the effective index of each respective mode of $\ell$ topological charge) and detuning the probe center frequency $\nu_0$ by a specific amount. A change in the optical path difference $\Delta(\Delta n(z) \cdot dz)$ can therefore be adequately compensated by a change in center frequency, such that
\begin{equation}
    \frac{\Delta \nu_0}{\nu_0} = \frac{\Delta(\Delta n(z) \cdot dz)}{(\Delta n(z) \cdot dz)}.
    \label{eq:opd}
\end{equation}

A simple strategy for interrogation of all sensing points, then, consists of probing the fiber under test (FUT) with multiple center frequencies and reconstructing the frequency response of each effective interferometer formed at every position in the fiber. A perceived shift in the frequency response will therefore be proportional to the optical path difference, according to the relation given in equation \ref{eq:opd}. Strain and temperature can then be inferred from well known fiber coefficients that take into account the total length change, elasto-optic effect, thermal expansion and thermo-optic effect \cite{Koyamada2009}.

\section{Experimental demonstration}

\begin{figure}[!ht]
\centering
\includegraphics[width=\linewidth]{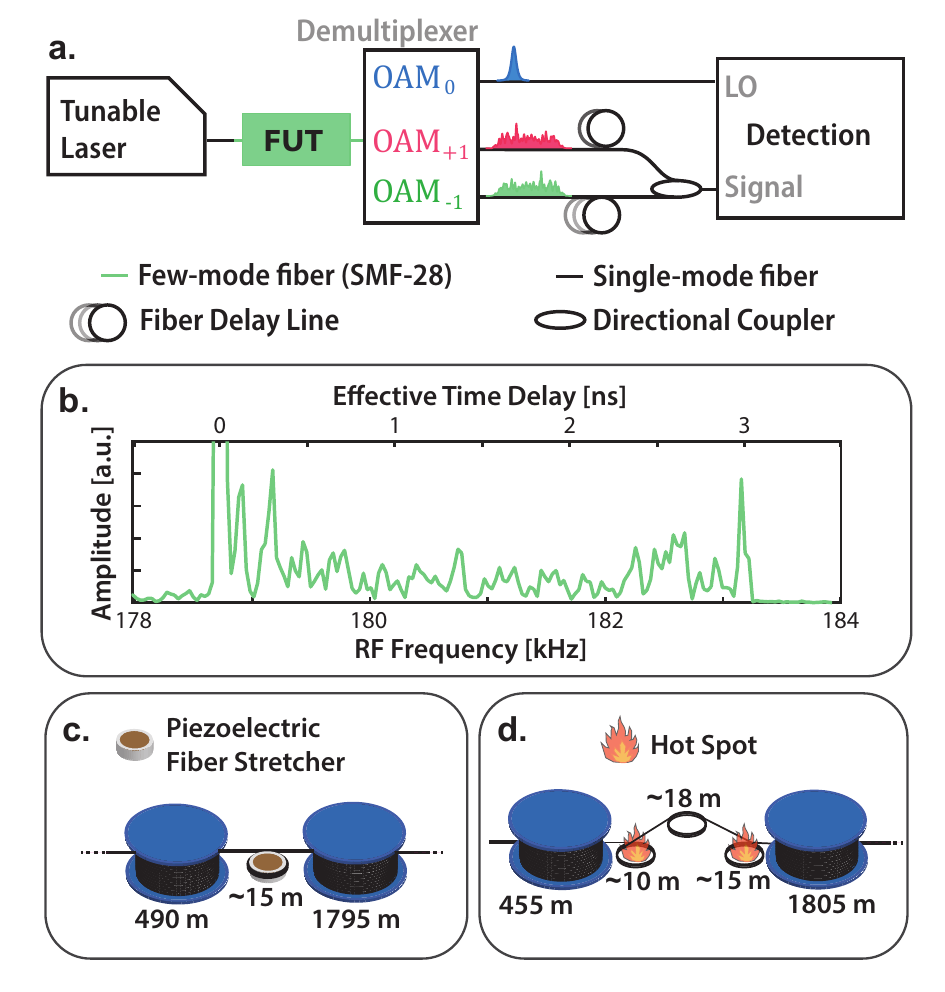}
\caption{\textbf{a.} Simplified depiction of the interrogation setup. The complete setup is described in the methods section (figure \ref{fig:fullsetup}).
\textbf{b.} Recovered impulse response obtained from one sub-sweep, (for a single mode and polarization) using swept-wavelength interferometry. Amplitude is normalized to the initial peak at time delay 0, occuring from imperfect demultiplexing at the output.
\textbf{c.} FUT configuration for the single-point strain measurement and \textbf{d.} for the multipoint temperature measurement.}
\label{fig:simplified_setup}
\end{figure}

To experimentally demonstrate our proposed method, we depart from our previous simplified conceptual description. The main difference is the use of a continuous-wave frequency swept input instead of a pulse for interrogation. This type of interrogation is commonly seen in other distributed sensing schemes \cite{s18041072-PhaseOTDROFDR}, and generally enables much improved spatial resolutions compared to time-domain (pulsed) implementations. We note that high resolution swept-wavelength interferometry (SWI) \cite{moore2011advances} techniques in backscattering methods are known to struggle in probing long lengths of fiber due to the limited coherence length of available sources \cite{rangeOFDR}. However, this trade-off is massively relaxed in our design by having the local oscillator (OAM\textsubscript{0} output) and measurement path (OAM\textsubscript{ $\pm$ 1} output) travel through the same fiber, being therefore automatically optical path-length-matched.

The use of a frequency-domain interrogation method has two main advantages: first, it facilitates the use of the ballistic OAM\textsubscript{0} output as the local oscillator, since the self-heterodyne and time-to-frequency mapping nature of the measurement enables the local oscillator to be conveniently spectrally separated (upconverted in the frequency domain) from the measurement outputs by the simple introduction of a fiber delay; and second, the improved spatial resolution achievable from SWI facilitates the demonstration of proof-of-principle in a benchtop experiment, using shorter perturbation lengths. This is particularly relevant considering the intrinsic penalty to spatial resolution by our proposed technique versus backscattering methods.

The maximum achievable spatial resolution by our method ($\zeta_{max}$) is proportional to the total bandwidth spanned by the sweep and the DGD, $\zeta_{max} \propto \frac{1}{B_{max} \cdot DGD}$.

For our specific implementation, however, each acquired time-series is divided into several sub-series, each corresponding to the output of a sub-sweep of bandwidth $B_{sub} < B_{max}$ and a slightly different laser center frequency $\nu_0$. This enables the reconstruction the frequency response of the fiber from a single swept acquisition, by simultaneously probing the fiber with multiple center frequencies at the cost of spatial resolution.

The resulting spatial resolution $\zeta$ is therefore calculated by
\begin{equation}
 \zeta = \frac{1}{B_{sub} \cdot DGD},
 \label{eqn:SR}
\end{equation}
and can be determined in post-processing by choosing the total bandwidth (or time) of each sub-sweep that the acquired portion of the scan is sliced into. This interrogation method also leads to an inverse proportionality between measurand resolution, spatial resolution and total bandwidth $B_{max}$, due to the limits of estimation accuracy of the frequency detuning \cite{picostrain} (see Appendix \ref{app:crlb}).

For our experimental demonstrations, we measured a 2300 meter long step-index SMF-28 fiber (carrying 2 mode groups at 1064 nm). The FUT configurations used in each experiment are represented in figure \ref{fig:simplified_setup}.

\subsection*{Multipoint Temperature Measurements}

To first evaluate the potential for distributed sensing and validate the principle for localization and discrimination of multiple measurements within the fiber, we performed a multipoint sensing measurement by heating two positions in the fiber. We generated two hotspots by coiling two sections of fiber ($\sim$10 m long and $\sim$15 m long) separated by more than one spatial resolution (figure \ref{fig:simplified_setup}). Room temperature was measured to be approximately 22 $ ^\circ C $.

Each of the fiber coils were heated by hovering a warm object ($\sim$ 35 $ ^\circ C $) close to the coils for about 1 minute without touching the fiber, and then removing it and allowing that hot spot to cool down. The spatial resolution was calculated to be 16.1 m according to equation \ref{eqn:SR}. 

\begin{figure}[!htb]
\centering
\includegraphics[width=0.9\linewidth]{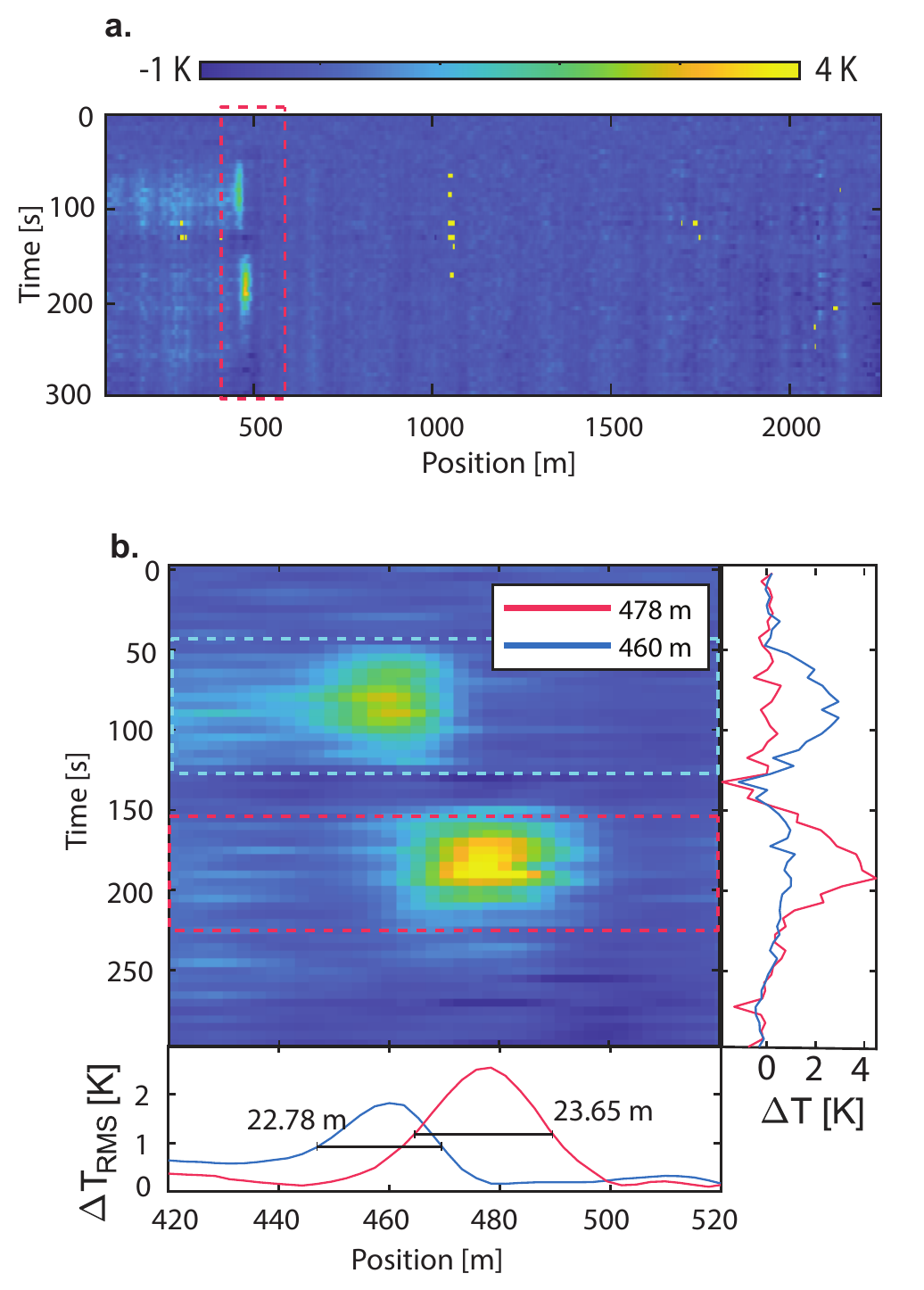}
\caption{Multipoint temperature measurement. \textbf{a.} The full fiber temperature profile. \textbf{b.} Close up of the perturbation region. Dashed regions mark sections to calculate the root-mean-square temperature shift to observe the spatial resolution (bottom subplot). Right subplot shows the temperature measurement at the point of highest perturbation amplitude for each hotspot over time.}
\label{fig:temperature}
\end{figure}

The results are depicted in figure  \ref{fig:temperature}. The spatial separation between both perturbations is clearly evidenced by computing the RMS temperature shift in the dashed areas, and plotting them in the bottom part of the figure. Notably, we see that the spatial full-width at half maximum for each perturbation is observed to be 22.78 m for the $\sim$10 m long hot spot and 23.65 for the $\sim$15 m long one. These results are reasonably consistent with the estimated lengths for the coils and the calculated spatial resolutions. Nonetheless, they seem to suggest some worsening of spatial resolution, which is expected to occur due to fluctuations in the laser sweep rate over each acquisition.

The temperature is calculated from the apparent effective frequency shift using standard coefficients used for telecommunication step-index fibers \cite{Koyamada2009}

\begin{equation}
    \Delta T \approx -\frac{1}{6.92 \times 10^{-6}} \frac{\Delta \nu_0}{\nu_0}.
    \label{eq:temperature}
\end{equation}

We notice the appearance of some residual crosstalk between spatial channels at positions prior to the measurement. While crosstalk effects have been observed for other Rayleigh-based distributed sensing methods, there are clear distinctions when compared to our approach. First, spatial crosstalk normally affects subsequent positions in backscattering-based technologies, and can be calculated from the amplitude of the perturbation that induces it \cite{Crosstalk}. This is contrasted with what we observe in the transmission setup: the crosstalk affects positions prior to the point of perturbation, and does not seem to scale predictably with the perturbation that induces it. This suggests that the origin might be an indirect effect, onset by changes to the strong coupling between the two degenerate OAM\textsubscript{ $\pm$ 1} modes. One possible explanation in this case may be a change in coupling strength from a combination of the fiber coiling and thermal effects. This is supported by the fact that the shorter coil had approximately half of the coiling radius as the longer coil (leading to stronger crosstalk). This may not occur in fiber installations that are not substantially bent or coiled, and may be avoided altogether through the use of a nondegenerate higher order mode for interrogation (thus avoiding any strong coupling effects), in fibers carrying a higher number of modes.

\subsection*{Strain Measurements}

\begin{figure}[!htb]
\centering
\includegraphics[width=\linewidth]{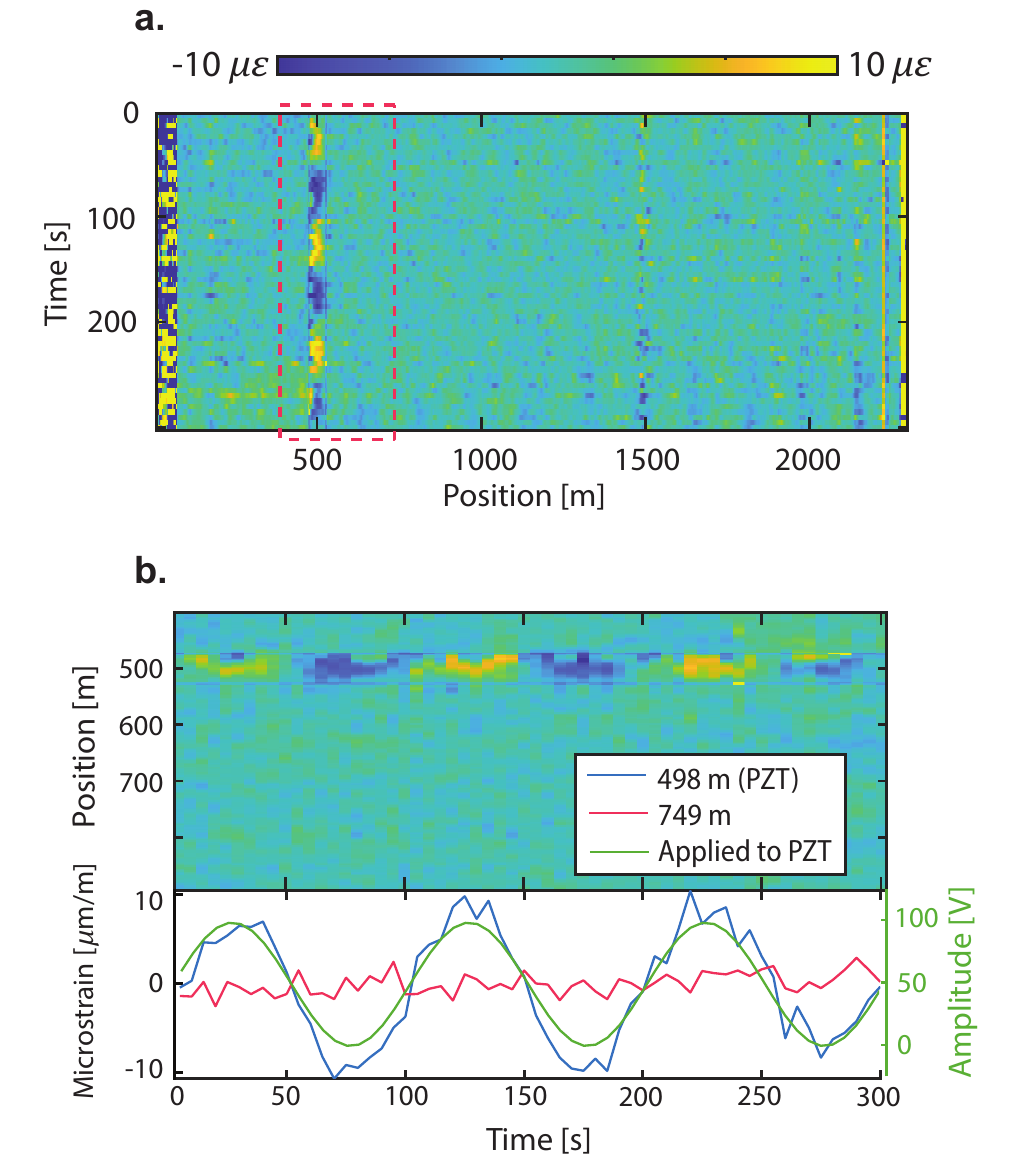}
\caption{Strain measurement. \textbf{a.} The full fiber strain profile. \textbf{b.} Close up of the perturbation region. The subplot shows the strain signal at PZT section of fiber (blue) and of an undisturbed position in the fiber (red), for comparison. Green shows the waveform applied to the piezoelectric fiber stretcher.}
\label{fig:strain}
\end{figure}

To evaluate the potential for strain measurements and assess the linearity of our interrogation process, we coiled a roughly 15 m long section of the fiber around a piezoelectric cylinder, at meter 500. A slow sinusoidal oscillation with 100 s period was applied to the fiber stretcher, with 100 V amplitude ($\sim$ 100 n$\varepsilon$/V, according to specifications). The strain distribution was recovered over 300 seconds, and is represented in figure \ref{fig:strain}. The acquired effective frequency shift was converted to strain through the following relation \cite{Koyamada2009}

\begin{equation}
    \varepsilon \approx -\frac{1}{0.78} \frac{\Delta \nu_0}{\nu_0}
    \label{eq:strain}
\end{equation}

Once again, the spatial resolution was selected to be 16.1 m, which was found to maximize the strain SNR for the recovered perturbation. The amplitude of the measured strain sine wave was found to be 16.4 $\mu \varepsilon$, and the strain resolution (computed as the average of the standard deviations of all points in an undisturbed section of fiber, from meter 700 to 1600) was measured as 1.2 $\mu \varepsilon$.

% \section{Methods}

\subsection*{Experimental Details}
\begin{figure*}[htbp]
\centering
\includegraphics[width=\linewidth]{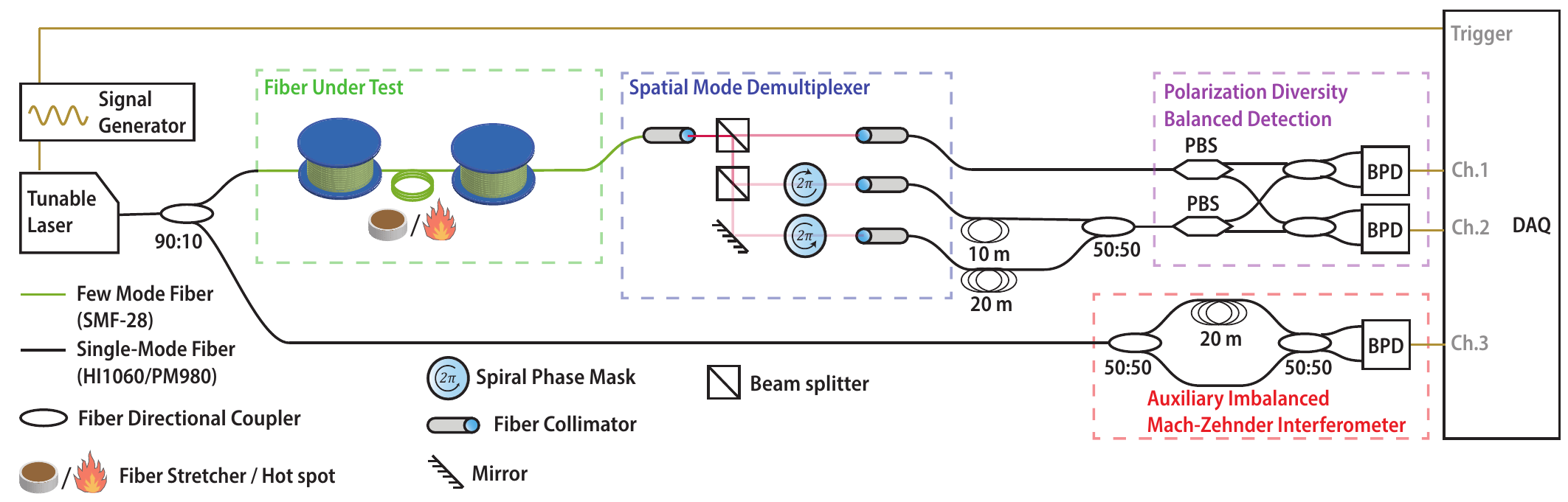}
\caption{Experimental setup. PBS: Polarization Beam Splitter; BPD: Balanced Photodetector; DAQ: Data Acquisition.}
\label{fig:fullsetup}
\end{figure*}

The full schematic for our setup is depicted in figure \ref{fig:fullsetup}. The laser source used was a Toptica CTL 1050, operating at center wavelength 1064 nm and swept by driving the internal stepper motor with a 0.2 Hz sine wave. A portion (10\%) of the laser output power is diverted into an imbalanced Mach-Zehnder interferometer used for compensation of sweep nonlinearity. The remaining 90\% ($\sim$ 12 dBm) is launched into the FUT. The laser is swept at an average rate of 1.63 THz/s (6.19 nm/s) over each acquisition (0.83 s around the point of highest linearity of the positive slope of the sinusoidal modulation).

The FUT consisted of 2.3 km of SMF-28 fiber carrying three modes over the measurement wavelength range (2 non-degenerate mode groups, OAM\textsubscript{0} and OAM\textsubscript{$\pm$ 1}), with DGD measured to be 1.23 ps/m. The differential mode group delay (DGD) of the fiber is inferred through previous knowledge of the length of fiber and by deliberately coupling a combination of OAM\textsubscript{$\pm$ 1} and OAM\textsubscript{0} light at the input, and observing the total delay between the high power peaks resulting from ballistic light at the output. 

At the fiber input, only the OAM\textsubscript{0} mode is excited. At the output, the three spatial modes are separated in a free-space section by collimating the fiber output and splitting it into 3 paths. One of the paths is immediately coupled into a single-mode HI1060 fiber, without undergoing any mode conversion, while the other two are sent through spiral phase masks (which add/subtract 1 topological charge) before being coupled to a single-mode HI1060 fiber. The spiral phase masks function as spatial mode converters, while the single-mode fibers act as spatial rejection filters that only accept the portion of light with 0 topological charge. The OAM\textsubscript{0} output is then used as the local oscillator of a polarization diversity balanced detection scheme. Each of the OAM\textsubscript{+1}/OAM\textsubscript{-1} outputs is delayed by approximately 10/20 meters of fiber (respectively), and then combined through a 50:50 fiber directional coupler. The addition of this delay upconverts each optical trace beatnote resulting from the heterodyne measurement, enabling both optical traces to be recovered in a single measurement \cite{Fontaine2015-charact,Rommel2017-charact}.

In order to achieve time-to-frequency mapping with a highly nonlinear frequency sweep, we calibrate the sweep rate of the laser using the auxiliary Mach-Zehnder interferometer (see figure \ref{fig:fullsetup}).  This allows us to correct the effects of nonlinearity by resampling the recovered time series with the instantaneous phase of the auxiliary interferometer signal $s_{aux}$, obtained by
\begin{equation}
    \phi_{inst} (t) = \measuredangle \mathcal{H} \{ s_{aux} (t) \},
\end{equation}
where $\mathcal{H}$ is the Hilbert transform. The OAM\textsubscript{+1}/OAM\textsubscript{-1} output time-series are then resampled with $\phi_{inst} (t-\tau_{FUT})$, where $\tau_{FUT}$ is the total delay accumulated by propagation through the FUT and fibers on the detection setup. 

After correction of sweep nonlinearity, the frequency responses of all sensing channels in the fiber can be acquired in a single shot by partitioning each acquired time-series (of length $t_{acq}$) into $N_s$ sub-sections of length $t_{sub} < t_{acq}$. Each of these sub-sections yields a different optical trace, equivalent to probing the fiber with a sweep of different center frequency and covering a narrower bandwidth ($B_{sub} \approx \langle \gamma(t) \rangle \times t_{sub}$, for an average laser sweep-rate $\langle \gamma(t) \rangle$). 

All optical traces obtained in this way (from Fourier transforming each sub-sweep) are then interpolated to a predetermined number of samples corresponding to the number of sensing channels we wish to record, therefore correcting any fluctuation of the optical trace width due to the laser sweep rate differences between sub-sections or acquisitions.

The interpolated optical traces obtained from all sub-sweeps are then stored as the columns of a matrix: each row stores the frequency response for each individual sensing channel (sensing position), as long as the Nyquist sampling criterion is satisfied

\begin{equation}
    \delta \nu < \frac{B_{sub}}{2},
\end{equation}
where $\delta \nu$ is the center frequency difference between each successive sub-sweep, and can be estimated by $\delta \nu \approx \langle \gamma(t) \rangle \times \Delta t_{sub}$, $\Delta t_{sub}$ being the time interval between adjacent sub-sections of the acquired time-series.

\begin{figure}[htbp]
\centering
\includegraphics[width=\linewidth]{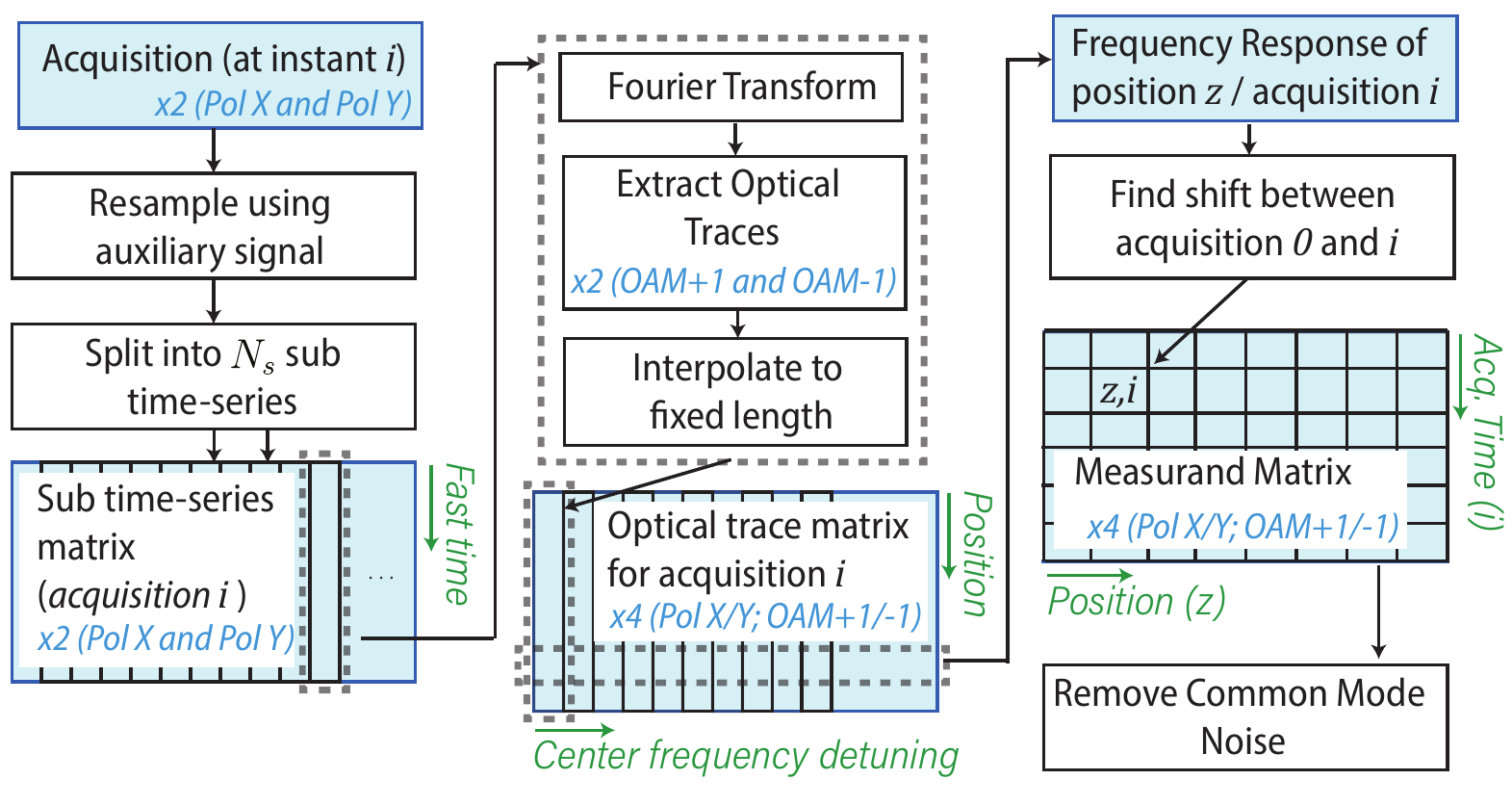}
\caption{Signal processing stack for an acquisition. Light blue marks signals or matrices and white marks processing steps.}
\label{fig:processingstack}
\end{figure}

This processing is repeated for all of the four FUT outputs (each pair of output modes and polarizations). Each of these frequency response matrices are independently processed to produce an estimation of the measurand, and then combined by averaging. The full processing stack can be visualized in figure \ref{fig:processingstack}. 

Estimation of measurand amplitude is accomplished by observing the detuning of the frequency responses of each sensing channel. The frequency detuning estimation between the m-th and r-th (reference) acquisitions is accomplished through the generalized cross-correlation algorithm

\begin{equation}
  \Delta \nu^{eff}_m(z_i) = \mbox{arg max} \{R^{(z_i)}_{m,r}(\Delta \nu)\},
\end{equation}
 where $\Delta \nu^{eff}_m(z_i)$ is the effective frequency detuning, proportional to the applied local perturbation to the fiber, and $R^{(z_i)}_{m,r}(\Delta \nu)$ is the cross-correlation between the frequency responses acquired at the instant $m$ and $r$, for the $z_i$-th measurement point. Subsample accuracy is achieved through parabolic fitting using the three points surrounding the maximum of the cross-correlation.

After recovering the full measurand profile, drifts or fluctuations of the center frequency of the laser are  corrected by removing the mean strain/temperature obtained along an unperturbed section of fiber, since they manifest as a spatially correlated common-mode noise component (from meter 700 to 1600 in our experiments) \cite{rosarioCMN}.

The perturbation for the strain experiment was accomplished by having approximately 15 m of fiber coiled around a piezoelectric cylinder, used as a fiber stretcher. The perturbation was a 100 V amplitude sine wave with 100 second period, with a 50 V offset in order to pre-stretch the fiber and prevent strain non-uniformities. The perturbation for the temperature measurements was achieved by hovering a warm object ($\sim$ 35 $^o$C) about 1 cm above the coiled fiber constituting the hotspot. At every acquisition, the signals are directly sent to a computer for storage and processing. 

\section{Discussion}
In this work, we introduced and demonstrated a new method to perform distributed sensing of common physical parameters by exploring the weak coupling between spatial modes carried in optical fibers, relying exclusively on unidirectional propagation in the fiber. 

As a proof-of-concept, we performed two sensing demonstrations using standard step-index SMF-28 working in few-mode operation. We successfully localized and demonstrated linear measurements of strain/temperature with inferred measurand resolutions of 1.2 $\mu \varepsilon$ (0.135 K, in equivalent temperature), and spatial resolutions in the tens of meters, at acquisition rates of 0.2 Hz. While these values are not representative of the ultimate performance limits of the technique and there is ample room for optimization, they serve as strong evidence for the future potential of this sensing principle. In particular, optical SNR is limited by the imperfect correction of the laser sweep nonlinearity and imperfect spatial mode demultiplexing at the fiber output, and the acquisition rate is limited by the total time required to ensure an approximately linear sweep from the laser source. Also, despite the slow acquisition rates, we note that this type of interrogation in fact relaxes the fundamental acquisition rate limit of common backscattering implementations, which require the minimum of full roundtrip time for the whole length of interrogated fiber between acquisitions. Conversely, our method benefits from a much narrower recovered optical trace such that this condition is massively relaxed, and multiple probe pulses can simultaneously coexist in the same FUT.

We also highlight the important distinction of our method from single-mode implementations, due to the potential to output multiple optical traces at every measurement. As such, our method may enjoy other benefits commonly mentioned for multimode-based fiber sensors \cite{9209959-sdmsensrev}, onset from the ability to access multiple mode outputs with different optical properties. In the presented work, we demonstrated the simplest case, with the minimum possible number of modes in a few-mode circularly symmetric fiber. However, our method generalizes to fibers carrying a higher number of modes or coupled-core multicore fibers, so long as the conditions of having access to a pair of weakly coupled modes (or supermodes) with differing group velocities is fulfilled. The ability to access multiple optical trace outputs also opens new processing possibilities, which may range from simple averaging of incoherent sources of noise (as done in this work), to more advanced processing schemes aimed at preventing the onset of anomalous estimations from cross-correlation \cite{LiLargeErrors}, and thus increasing the sensing dynamic while mitigating the consequent accumulation of 1/f noise \cite{Bhatta:19}.

We also note that our simple design can be readily adapted for different sensing paradigms, such as previously reported distributed transverse stress by the addition of simple processing steps, such as averaging of the optical traces acquired from each subsweep to produce an incoherent measurement of the coupling strength envelope \cite{Jia2021-Pressure}. Notably, since both methodologies (coherent interrogation for optical path measurements, and incoherent interrogation for coupling strength measurements) differ only in the post-processing, they can in principle be implemented simultaneously.

Despite the remaining optimization efforts required for a field demonstration, we stress the potential of this design for future seismic sensing, in space-division multiplexed telecommunication links consisting of weakly coupled fiber links. Further investigation should provide answers on the compatibility and total interrogation range achievable with this technique in telecommunication-grade few-mode fibers (using in-line amplifiers), and what performances can be expected when using kilometer-length spatial resolutions. With the aim of ultralong range, kilometer-length spatial resolutions, and strain measurements in mind, a potential roadmap for future designs based on this technology may include the use of a mixed pulsed/swept approach, analogous to some works described in the Rayleigh-backscattering based sensing literature \cite{Liu2015a-Range,picostrain}. This would entail limiting the total interrogated frequency range to a single wavelength channel, spanning multiple weakly coupled spatial channels. Our method can be adapted to such an implementation with only a few alterations to the hardware and processing scheme, by replacing the single-sweep approach into a multi-shot interrogation where the center frequency of each pulse is slowly modulated. This type of interrogation would benefit from the potential to co-propagate multiple pulses in the same fiber due to the transmission-based nature of the technique.

\section*{Acknowledgment}
The authors thank Optiphase/Jeff Bush for providing the piezoelectric fiber stretcher used in the strain experiments.

\appendices

\section{Measurand resolution vs. Spatial resolution}
\label{app:crlb}
\begin{figure}[h]
\centering
\includegraphics[width=0.8\linewidth]{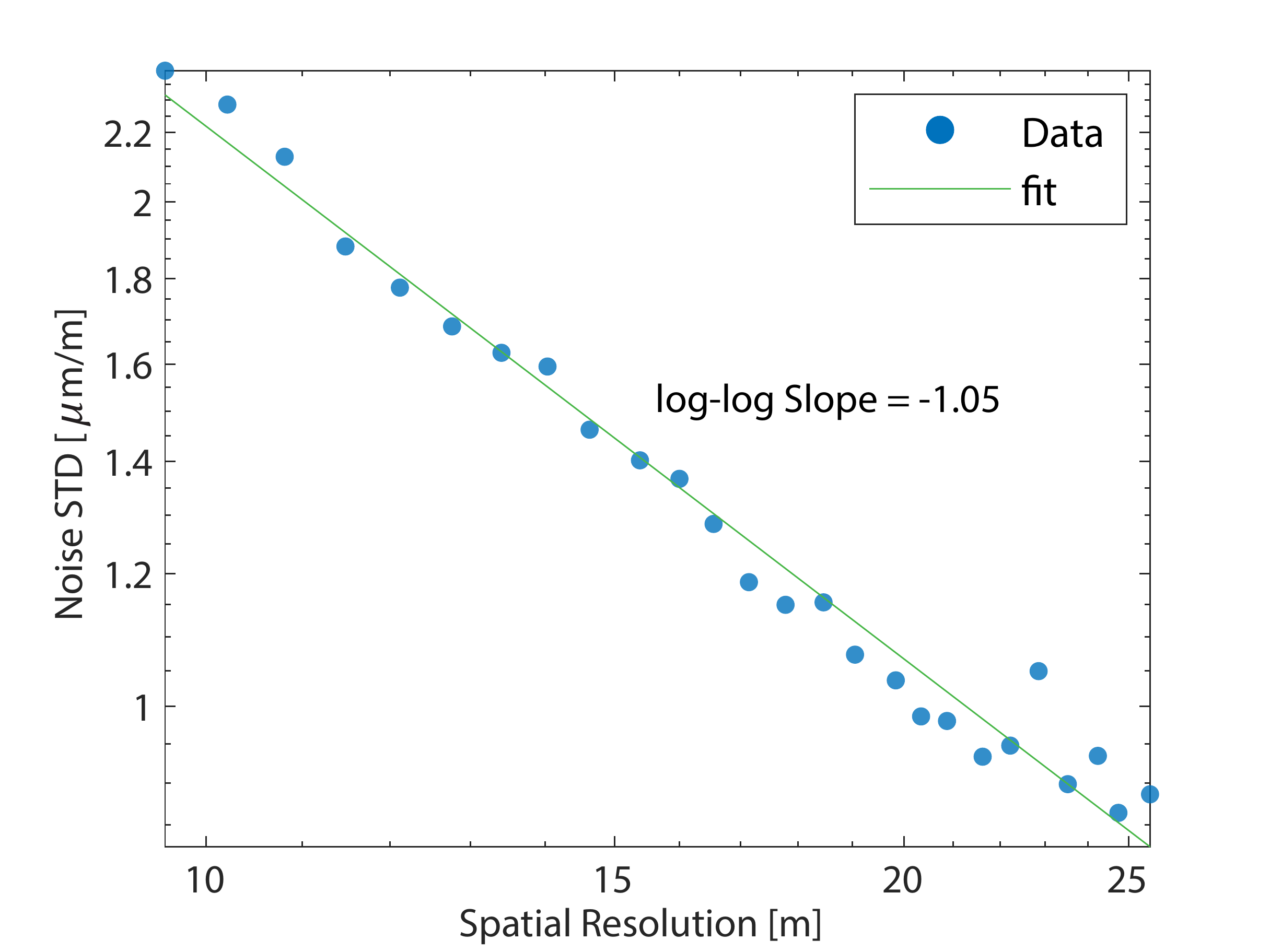}
\caption{Noise standard deviation (calculated in an unperturbed section of the fiber) as a function of spatial resolution. The log-log slope suggests a relationship of $\sigma \propto (SR)^{-1.05}$, which is close to our predicted CRLB scaling.}
\label{fig:crlb}
\end{figure}

The finite bandwidth covered by the total frequency sweep, as well as the presence of additive noise in the optical traces, lead to a fundamental limit to the estimation accuracy of a frequency detuning.

The accuracy limit for a frequency detuning has been studied for the analogous problem of time-delay estimation in distributed sensing systems \cite{picostrain}, and is determined by the Cramer-Rao Lower Bound (CRLB) of the estimation process, for the additive-noise limited case. Assuming a well conditioned signal for cross-correlation, the lower bound for a time-delay estimation measurement scales as follows, for a frequency-sweep based measurement \cite{picostrain}

\begin{equation}
    \sigma^2 \propto \frac{1}{SNR} \frac{1}{B_{max}} B_{sub}^3,
    \label{eqn:CRLBsweep}
\end{equation}
where $B_{max} = \gamma t$ ($t$ being the total acquisition time), and $B_{sub} = \frac{1}{DGD \cdot SR}$, $SR$ being the inferred spatial resolution of the system. Assuming linear scaling of SNR with total subsweep time ($t_{sub} = t \cdot \frac{B_{sub}}{B_{max}}$), we can write the SNR as $SNR = \frac{1}{C}\cdot (t \cdot \frac{1}{DGD \cdot SR \cdot \gamma t})$, $C$ being a constant factor with units of time. Equation \ref{eqn:CRLBsweep} can now be written as

\begin{equation}
    \sigma^2 \propto C \frac{1}{t} \left (\frac{1}{DGD \cdot SR} \right)  ^2,
    \label{eqn:CRLBsweeps}
\end{equation}
which highlights the inverse scaling between noise standard deviation and spatial resolution
\begin{equation}
    \sigma \propto SR^{-1},
    \label{eqn:scaling}
\end{equation}
We investigated this claim by doing a parameter sweep of the spatial resolution on our strain measurement data, and plotting the results in figure \ref{fig:crlb}. The standard deviation of noise was calculated as the average standard deviation in an unperturbed section of fiber, from meter 700 to 1600. While preliminary, this analysis suggests that the current strain resolution bottleneck is given by additive noise sources, and hints at the performance scaling.

%%
%

% Bibliography

\ifCLASSOPTIONcaptionsoff
  \newpage
\fi



%\bibliographystyle{IEEEtran}
%\begin{thebibliography}{1}
\bibliographystyle{IEEEtran}
\bibliography{IEEEabrv,sample}

\begin{thebibliography}{10}
\providecommand{\url}[1]{#1}
\csname url@samestyle\endcsname
\providecommand{\newblock}{\relax}
\providecommand{\bibinfo}[2]{#2}
\providecommand{\BIBentrySTDinterwordspacing}{\spaceskip=0pt\relax}
\providecommand{\BIBentryALTinterwordstretchfactor}{4}
\providecommand{\BIBentryALTinterwordspacing}{\spaceskip=\fontdimen2\font plus
\BIBentryALTinterwordstretchfactor\fontdimen3\font minus
  \fontdimen4\font\relax}
\providecommand{\BIBforeignlanguage}[2]{{%
\expandafter\ifx\csname l@#1\endcsname\relax
\typeout{** WARNING: IEEEtran.bst: No hyphenation pattern has been}%
\typeout{** loaded for the language `#1'. Using the pattern for}%
\typeout{** the default language instead.}%
\else
\language=\csname l@#1\endcsname
\fi
#2}}
\providecommand{\BIBdecl}{\relax}
\BIBdecl

\bibitem{he2021optical-review-das}
Z.~He and Q.~Liu, ``{Optical Fiber Distributed Acoustic Sensors: A Review},''
  \emph{Journal of Lightwave Technology}, vol.~39, no.~12, pp. 3671--3686,
  2021.

\bibitem{Lu2019-review}
P.~Lu, N.~Lalam, M.~Badar, B.~Liu, B.~T. Chorpening, M.~P. Buric, and P.~R.
  Ohodnicki, ``{Distributed optical fiber sensing: Review and perspective},''
  \emph{Applied Physics Reviews}, vol.~6, no.~4, 2019.

\bibitem{Bashan2018-Impedance}
G.~Bashan, H.~H. Diamandi, Y.~London, E.~Preter, and A.~Zadok,
  ``{Optomechanical time-domain reflectometry},'' \emph{Nature Communications},
  vol.~9, no.~1, p. 2991, 2018.

\bibitem{chow2018distributed-Impedance}
D.~M. Chow, Z.~Yang, M.~A. Soto, and L.~Th{\'{e}}venaz, ``{Distributed forward
  Brillouin sensor based on local light phase recovery},'' \emph{Nature
  communications}, vol.~9, no.~1, pp. 1--9, 2018.

\bibitem{Bashan2020-Impedance}
G.~Bashan, Y.~London, H.~{Hagai Diamandi}, and A.~Zadok, ``{Distributed
  cladding mode fiber-optic sensor},'' \emph{Optica}, vol.~7, no.~1, p.~85,
  2020.

\bibitem{Jia2021-Pressure}
J.~Jia, Y.~Yang, M.~Zuo, J.~Cui, Y.~Gao, J.~Yu, H.~Yu, Z.~R. Zhang, Z.~Chen,
  Y.~He, and J.~Li, ``{Distributed Transverse Stress Sensor Based on Mode
  Coupling in Weakly-Coupled FMF},'' \emph{IEEE Photonics Journal}, vol.~14,
  no.~1, pp. 1--1, 2021.

\bibitem{ZhangLi-Pressure}
L.~Zhang, Z.~Yang, L.~Szostkiewicz, K.~Markiewicz, S.~Mikhailov, T.~Geernaert,
  E.~Rochat, and L.~Thévenaz, ``Long-distance distributed pressure sensing
  based on frequency-scanned phase-sensitive optical time-domain
  reflectometry,'' \emph{Optics Express}, vol.~29, 06 2021.

\bibitem{Zhou2018-StrainTemp}
D.~Zhou, Y.~Dong, B.~Wang, C.~Pang, D.~Ba, H.~Zhang, Z.~Lu, H.~Li, and X.~Bao,
  ``{Single-shot BOTDA based on an optical chirp chain probe wave for
  distributed ultrafast measurement},'' \emph{Light: Science and Applications},
  vol.~7, no.~1, p.~32, 2018.

\bibitem{Soriano-Amat2021-StrainTemp}
M.~Soriano-Amat, H.~F. Martins, V.~Dur{\'{a}}n, L.~Costa, S.~Martin-Lopez,
  M.~Gonzalez-Herraez, and M.~R. Fern{\'{a}}ndez-Ruiz, ``{Time-expanded
  phase-sensitive optical time-domain reflectometry},'' \emph{Light: Science
  and Applications}, vol.~10, no.~1, 2021.

\bibitem{Rao2021-PhaseOTDROFDR}
Y.~Rao, Z.~Wang, H.~Wu, Z.~Ran, and B.~Han, ``{Recent Advances in
  Phase-Sensitive Optical Time Domain Reflectometry ($\varphi$-OTDR)},''
  \emph{Photonic Sensors}, vol.~11, no.~1, pp. 1--30, 2021.

\bibitem{Lior2021-GeoDas}
I.~Lior, A.~Sladen, D.~Rivet, J.~P. Ampuero, Y.~Hello, C.~Becerril, H.~F.
  Martins, P.~Lamare, C.~Jestin, S.~Tsagkli, and C.~Markou, ``{On the Detection
  Capabilities of Underwater Distributed Acoustic Sensing},'' \emph{Journal of
  Geophysical Research: Solid Earth}, vol. 126, no.~3, pp. 1--20, 2021.

\bibitem{Hammond2019-GeoDas}
J.~O. Hammond, R.~England, N.~Rawlinson, A.~Curtis, K.~Sigloch, N.~Harmon, and
  B.~Baptie, ``{The future of passive seismic acquisition},'' \emph{Astronomy
  and Geophysics}, vol.~60, no.~2, pp. 37--42, 2019.

\bibitem{Williams2019-GeoDas}
E.~F. Williams, M.~R. Fern{\'{a}}ndez-Ruiz, R.~Magalhaes, R.~Vanthillo,
  Z.~Zhan, M.~Gonz{\'{a}}lez-Herr{\'{a}}ez, and H.~F. Martins, ``{Distributed
  sensing of microseisms and teleseisms with submarine dark fibers},''
  \emph{Nature Communications}, vol.~10, no.~1, pp. 1--11, 2019.

\bibitem{Fernandez-Ruiz2021-GeoDas}
M.~R. Fernandez-Ruiz, H.~F. Martins, E.~Williams, C.~Becerril, R.~Magalhaes,
  L.~D. Costa, S.~Martin-Lopez, S.~Jia, Z.~Zhan, and M.~Gonzalez-Herraez,
  ``{Seismic Monitoring with Distributed Acoustic Sensing from the Near-surface
  to the Deep Oceans},'' \emph{Journal of Lightwave Technology}, pp. 1--1,
  2021.

\bibitem{Martins:19-GeoDasDSP}
H.~F. Martins, M.~R. Fern{\'{a}}ndez-Ruiz, L.~Costa, E.~Williams, Z.~Zhan,
  S.~Martin-Lopez, and M.~Gonzalez-Herraez, ``{Monitoring of remote seismic
  events in metropolitan area fibers using distributed acoustic sensing (DAS)
  and spatiotemporal signal processing},'' in \emph{Optical Fiber Communication
  Conference (OFC) 2019}.\hskip 1em plus 0.5em minus 0.4em\relax Optica
  Publishing Group, 2019, p. M2J.1.

\bibitem{Nuno:21-Range}
J.~Nu{\~{n}}o, S.~Martin-Lopez, J.~D. Ania-Casta{\~{n}}{\'{o}}n,
  M.~Gonzalez-Herraez, and H.~F. Martins, ``{Virtual transparency in
  $\phi$-OTDR using second order Raman amplification and pump modulation},''
  \emph{Opt. Express}, vol.~29, no.~22, pp. 35\,725--35\,734, oct 2021.

\bibitem{Liu2015a-Range}
Q.~Liu, X.~Fan, and Z.~He, ``{Time-gated digital optical frequency domain
  reflectometry with 1.6-m spatial resolution over entire 110-km range},''
  \emph{Optics Express}, vol.~23, no.~20, pp. 3323--3328, 2015.

\bibitem{Suetsugu2014-OceanbottomTraditionalInstruments}
D.~Suetsugu and H.~Shiobara, ``{Broadband ocean-bottom seismology},''
  \emph{Annual Review of Earth and Planetary Sciences}, vol.~42, pp. 27--43,
  2014.

\bibitem{MarraScience-OceanFiberSensing}
G.~Marra, C.~Clivati, R.~Luckett, A.~Tampellini, J.~Kronj{\"{a}}ger, L.~Wright,
  A.~Mura, F.~Levi, S.~Robinson, A.~Xuereb, B.~Baptie, and D.~Calonico,
  ``{Ultrastable laser interferometry for earthquake detection with terrestrial
  and submarine cables},'' \emph{Science}, vol. 361, no. 6401, pp. 486--490,
  2018.

\bibitem{ZhanScience-OceanFiberSensing}
Z.~Zhan, M.~Cantono, V.~Kamalov, A.~Mecozzi, R.~M{\"{u}}ller, S.~Yin, and J.~C.
  Castellanos, ``{Optical polarization-based seismic and water wave sensing on
  transoceanic cables},'' \emph{Science}, vol. 371, no. 6532, pp. 931--936,
  2021.

\bibitem{Mecozzi:21-OceanFiberSensing}
A.~Mecozzi, M.~Cantono, J.~C. Castellanos, V.~Kamalov, R.~Muller, and Z.~Zhan,
  ``{Polarization sensing using submarine optical cables},'' \emph{Optica},
  vol.~8, no.~6, pp. 788--795, jun 2021.

\bibitem{Jia:21-Coexistence}
Z.~Jia, L.~A. Campos, M.~Xu, H.~Zhang, M.~Gonzalez-Herraez, H.~F. Martins, and
  Z.~Zhan, ``{Experimental Coexistence Investigation of Distributed Acoustic
  Sensing and Coherent Communication Systems},'' in \emph{Optical Fiber
  Communication Conference (OFC) 2021}.\hskip 1em plus 0.5em minus 0.4em\relax
  Optica Publishing Group, 2021, p. Th4F.4.

\bibitem{Winzer2020TransmissionSC-SDMTechno}
P.~J. Winzer, ``Transmission system capacity scaling through space-division
  multiplexing: a techno-economic perspective,'' 2020.

\bibitem{Puttnam:21-SDM}
B.~J. Puttnam, G.~Rademacher, and R.~S. Lu{\'{i}}s, ``{Space-division
  multiplexing for optical fiber communications},'' \emph{Optica}, vol.~8,
  no.~9, pp. 1186--1203, sep 2021.

\bibitem{8004172-FMF}
K.-i. Kitayama and N.-P. Diamantopoulos, ``{Few-Mode Optical Fibers: Original
  Motivation and Recent Progress},'' \emph{IEEE Communications Magazine},
  vol.~55, no.~8, pp. 163--169, 2017.

\bibitem{Saitoh:16-mcf}
K.~Saitoh and S.~Matsuo, ``{Multicore fiber technology},'' \emph{Journal of
  Lightwave Technology}, vol.~34, no.~1, pp. 55--66, jan 2016.

\bibitem{dar2018cost-SDMTechno}
R.~Dar, P.~J. Winzer, A.~R. Chraplyvy, S.~Zsigmond, K.-Y. Huang, H.~Fevrier,
  and S.~Grubb, ``{Cost-optimized submarine cables using massive spatial
  parallelism},'' \emph{Journal of Lightwave Technology}, vol.~36, no.~18, pp.
  3855--3865, 2018.

\bibitem{Li:15-sdmsensrev}
A.~Li, Y.~Wang, Q.~Hu, and W.~Shieh, ``{Few-mode fiber based optical
  sensors},'' \emph{Opt. Express}, vol.~23, no.~2, pp. 1139--1150, jan 2015.

\bibitem{9209959-sdmsensrev}
I.~Ashry, Y.~Mao, A.~Trichili, B.~Wang, T.~K. Ng, M.-S. Alouini, and B.~S. Ooi,
  ``{A Review of Using Few-Mode Fibers for Optical Sensing},'' \emph{IEEE
  Access}, vol.~8, pp. 179\,592--179\,605, 2020.

\bibitem{Zhao:21-sdmsensrev}
Z.~Zhao and M.~Tang, ``{Distributed fiber sensing using SDM fibers},'' in
  \emph{26th Optoelectronics and Communications Conference}.\hskip 1em plus
  0.5em minus 0.4em\relax Optica Publishing Group, 2021, p. W4D.1.

\bibitem{kim2021recent-sdmsensrev}
Y.~H. Kim and K.~Y. Song, ``{Recent Progress in Distributed Brillouin Sensors
  Based on Few-Mode Optical Fibers},'' \emph{Sensors}, vol.~21, no.~6, p. 2168,
  2021.

\bibitem{Li2015FewmodeFM}
A.~Li, Y.~Wang, J.~Fang, M.-J. Li, B.~Y. Kim, and W.~Shieh, ``{Few-mode fiber
  multi-parameter sensor with distributed temperature and strain
  discrimination.}'' \emph{Optics letters}, vol. 40 7, pp. 1488--1491, 2015.

\bibitem{8839098}
Y.~Mao, I.~Ashry, M.~S. Alias, T.~K. Ng, F.~Hveding, M.~Arsalan, and B.~S. Ooi,
  ``{Investigating the Performance of a Few-Mode Fiber for Distributed Acoustic
  Sensing},'' \emph{IEEE Photonics Journal}, vol.~11, no.~5, pp. 1--10, 2019.

\bibitem{9502556}
Y.~Meng, C.~Fu, C.~Du, L.~Chen, H.~Zhong, P.~Li, B.~Xu, B.~Du, J.~He, and
  Y.~Wang, ``{Shape Sensing Using Two Outer Cores of Multicore Fiber and
  Optical Frequency Domain Reflectometer},'' \emph{Journal of Lightwave
  Technology}, vol.~39, no.~20, pp. 6624--6630, 2021.

\bibitem{8513845}
Z.~Zhao, M.~Tang, L.~Wang, N.~Guo, H.-Y. Tam, and C.~Lu, ``{Distributed
  Vibration Sensor Based on Space-Division Multiplexed Reflectometer and
  Interferometer in Multicore Fiber},'' \emph{Journal of Lightwave Technology},
  vol.~36, no.~24, pp. 5764--5772, 2018.

\bibitem{s18041072-PhaseOTDROFDR}
Z.~Ding, C.~Wang, K.~Liu, J.~Jiang, D.~Yang, G.~Pan, Z.~Pu, and T.~Liu,
  ``{Distributed Optical Fiber Sensors Based on Optical Frequency Domain
  Reflectometry: A review},'' \emph{Sensors}, vol.~18, no.~4, 2018.

\bibitem{Murray:22}
J.~B. Murray, A.~Cerjan, and B.~Redding, ``{Distributed Brillouin fiber laser
  sensor},'' \emph{Optica}, vol.~9, no.~1, pp. 80--87, jan 2022.

\bibitem{Kono2017-charact}
N.~Kono, F.~Ito, D.~Iida, and T.~Manabe, ``{Impulse response measurement of
  few-mode fiber systems by coherence-recovered linear optical sampling},''
  \emph{Journal of Lightwave Technology}, vol.~35, no.~20, pp. 4392--4398,
  2017.

\bibitem{Rommel2017-charact}
S.~Rommel, J.~M.~D. Mendinueta, W.~Klaus, J.~Sakaguchi, J.~J.~V. Olmos,
  Y.~Awaji, I.~T. Monroy, and N.~Wada, ``{Few-mode fiber, splice and SDM
  component characterization by spatially-diverse optical vector network
  analysis},'' \emph{Optics Express}, vol.~25, no.~19, p. 22347, 2017.

\bibitem{Fontaine2015-charact}
N.~K. Fontaine, ``{Characterization of space-division multiplexing fibers using
  swept-wavelength interferometry},'' \emph{Optical Fiber Communication
  Conference, OFC 2015}, vol.~1, no.~c, pp. 4--6, 2015.

\bibitem{Maruyama2017-charact}
R.~Maruyama, N.~Kuwaki, S.~Matsuo, and M.~Ohashi, ``{Relationship between Mode
  Coupling and Fiber Characteristics in Few-Mode Fibers Analyzed Using Impulse
  Response Measurements Technique},'' \emph{Journal of Lightwave Technology},
  vol.~35, no.~4, pp. 650--657, 2017.

\bibitem{Jia:22}
J.~Jia, J.~Cui, J.~Zhang, M.~Zuo, Y.~Gao, Z.~Chen, Y.~He, and J.~Li,
  ``Distributed vibration sensor based on mode coupling in weakly coupled
  few-mode fibers,'' \emph{Opt. Lett.}, vol.~47, no.~7, pp. 1717--1720, Apr
  2022.

\bibitem{moore2011advances}
E.~D. Moore, ``{Advances in swept-wavelength interferometry for precision
  measurements},'' Ph.D. dissertation, University of Colorado at Boulder, 2011.

\bibitem{Willner:15}
A.~E. Willner, H.~Huang, Y.~Yan, Y.~Ren, N.~Ahmed, G.~Xie, C.~Bao, L.~Li,
  Y.~Cao, Z.~Zhao, J.~Wang, M.~P.~J. Lavery, M.~Tur, S.~Ramachandran, A.~F.
  Molisch, N.~Ashrafi, and S.~Ashrafi, ``{Optical communications using orbital
  angular momentum beams},'' \emph{Adv. Opt. Photon.}, vol.~7, no.~1, pp.
  66--106, mar 2015.

\bibitem{Wang:16}
Z.~Wang, L.~Zhang, S.~Wang, N.~Xue, F.~Peng, M.~Fan, W.~Sun, X.~Qian, J.~Rao,
  and Y.~Rao, ``{Coherent $\varphi$-OTDR based on I/Q demodulation and homodyne
  detection},'' \emph{Opt. Express}, vol.~24, no.~2, pp. 853--858, jan 2016.

\bibitem{Lu:20}
X.~Lu, M.~A. Soto, L.~Zhang, and L.~Th{\'{e}}venaz, ``{Spectral Properties of
  the Signal in Phase-Sensitive Optical Time-Domain Reflectometry With Direct
  Detection},'' \emph{J. Lightwave Technol.}, vol.~38, no.~6, pp. 1513--1521,
  mar 2020.

\bibitem{Koyamada2009}
Y.~Koyamada, M.~Imahama, K.~Kubota, and K.~Hogari, ``{Fiber-optic distributed
  strain and temperature sensing with very high measurand resolution over long
  range using coherent OTDR},'' \emph{Journal of Lightwave Technology},
  vol.~27, no.~9, pp. 1142--1146, 2009.

\bibitem{rangeOFDR}
F.~Ito, X.~Fan, and Y.~Koshikiya, ``{Long-Range Coherent OFDR With Light Source
  Phase Noise Compensation},'' \emph{Journal of Lightwave Technology}, vol.~30,
  no.~8, pp. 1015--1024, 2012.

\bibitem{picostrain}
L.~Costa, H.~F. Martins, S.~Martin-Lopez, M.~R. Fernandez-Ruiz, and
  M.~Gonzalez-Herraez, ``{Fully Distributed Optical Fiber Strain Sensor with
  $10^{-12}$ $\varepsilon$/$\sqrt{}$Hz Sensitivity},'' \emph{Journal of
  Lightwave Technology}, vol.~37, no.~18, pp. 4487--4495, 2019.

\bibitem{Crosstalk}
L.~Marcon, M.~Soriano-Amat, R.~Veronese, A.~Garcia-Ruiz, M.~Calabrese,
  L.~Costa, M.~R. Fernandez-Ruiz, H.~F. Martins, L.~Palmieri, and
  M.~Gonzalez-Herraez, ``{Analysis of Disturbance-Induced “Virtual”
  Perturbations in Chirped Pulse $\phi$-OTDR},'' \emph{IEEE Photonics
  Technology Letters}, vol.~32, no.~3, pp. 158--161, 2020.

\bibitem{rosarioCMN}
M.~R. Fernández-Ruiz, J.~Pastor-Graells, H.~F. Martins, A.~Garcia-Ruiz,
  S.~Martin-Lopez, and M.~Gonzalez-Herraez, ``Laser phase-noise cancellation in
  chirped-pulse distributed acoustic sensors,'' \emph{Journal of Lightwave
  Technology}, vol.~36, no.~4, pp. 979--985, 2018.

\bibitem{LiLargeErrors}
L.~Zhang, L.~D. Costa, Z.~Yang, M.~A. Soto, M.~Gonzalez-Herráez, and
  L.~Thévenaz, ``Analysis and reduction of large errors in rayleigh-based
  distributed sensor,'' \emph{Journal of Lightwave Technology}, vol.~37,
  no.~18, pp. 4710--4719, 2019.

\bibitem{Bhatta:19}
H.~D. Bhatta, L.~Costa, A.~Garcia-Ruiz, M.~R. Fernandez-Ruiz, H.~F. Martins,
  M.~Tur, and M.~Gonzalez-Herraez, ``{Dynamic Measurements of 1000 Microstrains
  Using Chirped-Pulse Phase-Sensitive Optical Time-Domain Reflectometry},''
  \emph{J. Lightwave Technol.}, vol.~37, no.~18, pp. 4888--4895, Sep 2019.

\end{thebibliography}

%\end{thebibliography}
 
\end{document}